\documentclass{article}

    \PassOptionsToPackage{numbers, compress}{natbib}

\usepackage[final]{neurips_2021}

\usepackage{graphicx,subcaption}

\usepackage[utf8]{inputenc} 
\usepackage[T1]{fontenc}    
\usepackage{hyperref}       
\usepackage{url}            
\usepackage{booktabs}       
\usepackage{amsfonts}       
\usepackage{nicefrac}       
\usepackage{microtype}      
\usepackage{xcolor}         
\usepackage[linesnumbered,ruled,vlined]{algorithm2e}

\usepackage{amssymb}
\usepackage{latexsym}
\usepackage{amsmath,amssymb}
\usepackage{graphicx}
\usepackage{multirow}
\usepackage{adjustbox}
\usepackage{makecell}
\usepackage[nottoc]{tocbibind}

\newcommand{\add}[1] {\textcolor{black}{#1}} 

\long\def\comment#1{} 

\newcommand{\Xc}{\mathcal{X}}

\newcommand{\Yc}{\mathcal{Y}}

\newcommand{\Ed}{{{\mathbb E}}}

\newcommand{\beq}{\begin{equation}}
\newcommand{\eeq}{\end{equation}}
\newcommand{\beqa}{\begin{eqnarray}}
\newcommand{\eeqa}{\end{eqnarray}}

\title{Noise2Score: Tweedie's Approach to Self-Supervised Image Denoising  without Clean Images}

\author{%
Kwanyoung Kim$^{1}$ \quad \quad Jong Chul Ye$^{1,2,3}$\\
  \normalfont  $^1$ Department of Bio and Brain Engineering\\
  $^2$Kim Jaechul Graduate School of AI\\
    $^3$Deptartment of Mathematical Sciences\\
   Korea Advanced Institute of Science and Technology (KAIST) \\
\texttt{\{cubeyoung, jong.ye\}@kaist.ac.kr} \\
}

\begin{document}

\maketitle

\begin{abstract}
Recently, there has  been extensive research interest in training  deep networks to denoise images without clean reference.
However, the representative approaches such as Noise2Noise, Noise2Void, Stein's unbiased risk estimator (SURE), etc.  seem to differ from one another and it is difficult to find the coherent mathematical structure.
To address this, here we present a novel approach, called Noise2Score, which reveals a missing link in order to unite these seemingly different approaches.
Specifically, we  show that   image denoising  problems  without clean images
can be addressed by finding the mode of the posterior distribution and that the Tweedie's formula offers an explicit solution
through the score function (i.e. the gradient of loglikelihood).
Our method then uses the  recent finding that  the score function  can be stably estimated
from the noisy images using the amortized residual denoising autoencoder, the method of which 
is closely related to Noise2Noise or Nose2Void.
Our Noise2Score approach is so universal  that the same network training can be used to remove noises from images that are corrupted by any exponential family distributions and noise parameters.
Using extensive  experiments with Gaussian, Poisson, and Gamma noises,
we show  that  Noise2Score significantly outperforms the state-of-the-art self-supervised denoising methods in the benchmark data set such as (C)BSD68, Set12, and Kodak, etc. 

\end{abstract}

\section{Introduction}

Bayesian inference, which derives the posterior probability using a prior probability and a likelihood function for the observed data, has been an important tool in  statistics.  
This approach has been used extensively  for image denoising from early ages to the modern era of deep learning.
For example, in the recent unsupervised deep learning approach for image denoising using Stein's risk
estimate (SURE) \cite{soltanayev2018training}, the unknown Bayesian risk is replaced by the
SURE that can be calculated from the noisy measurement so that deep neural network training is performed by minimizing it.
Unfortunately, this method is sensitive to hyper-parameters, and the neural network must be retrained
if the underlying noise model varies \cite{stein1981estimation}.
On the other hand, there has been increased research interest in  image denoisers that can be trained by minimizing variants of empirical risks that are not associated with clean data.
Noise2Noise ~\cite{lehtinen2018noise2noise} was the first representative approach that does not require clean data. 
Unfortunately, multiple noisy versions of the same images are necessary for training.
To address this,  self-supervised learning approaches such as
Noise2Void~\cite{krull2019noise2void}, Noise2Self~\cite{xie2020noise2same}, etc.  have been developed in order to
use only a single  noisy image. 
This class of approaches, which  we will call Noise2X throughout the paper, are especially important for practical applications, where noiseless clean images or multiple noisy realization of the same image are difficult or impossible to collect.

Though SURE and Noise2X hold promise for practical applications,
 one of the fundamental questions is how these seemingly different approaches are related and whether there is a coherent mathematical theory that can be leveraged to further improve the performance.
Although we are aware of a preliminary prior work \cite{zhussip2019extending} attempting to unite a subset of them, there is largely a lack of a principled method of designing image denoiser without clean data.

In this article, we try to approach this open problem  from Bayesian statistics in a completely different route.
Instead of minimizing  different forms of empirical risks or SURE, 
one of the most important contributions of this paper is the discovery of the importance of classical result from Bayesian statistics  - the Tweedie's formula \cite{efron2011tweedie} which provide an explicit way of computing the posterior mean of canonical parameters from the noisy measurements corrupted with exponential family noises.
Specifically, we show that the Tweedie's formula provides a unified approach for image denoising from any exponential family noises  through
a score function (i.e. the gradient of the loglikelihood).
Therefore, the self-supervised image denoising problem without clean image
can be reduced to the problem of estimating the score function.

\begin{figure}[!t]
	\centering
	\includegraphics[width=0.9\linewidth]{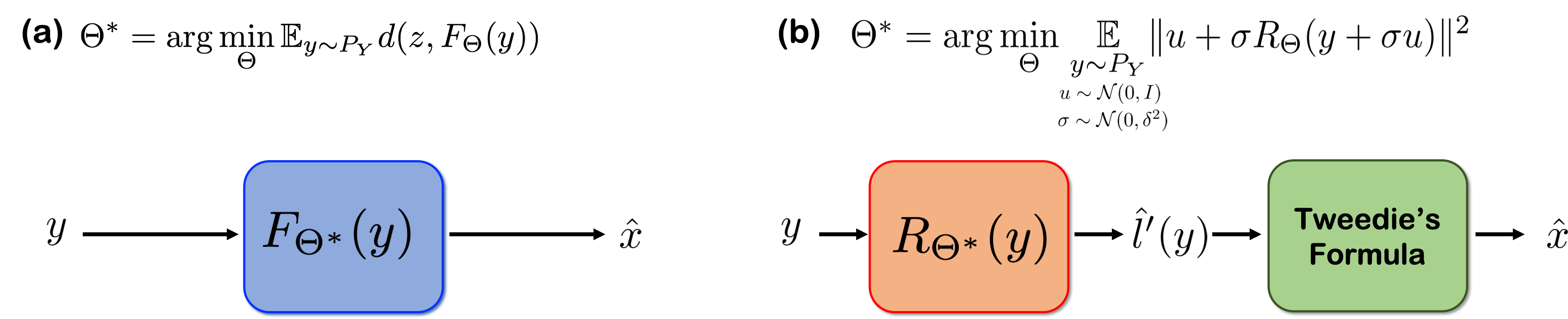}
	
	\caption{ Overall reconstruction flow of (a) supervised learning,  SURE and Noise2X, where  the target $z$ and the distance measure $d(\cdot,\cdot)$
	are uniquely determined by each algorithm, and (b) Noise2Score, where the first step is the estimation of the score function $\hat{l}'$ using neural network training, which is followed by 
	Tweedie's formula to obtain the final denoising result.}
	\label{fig:concept}
\end{figure}

This change of perspective has many important theoretical implications as well as the flexibility with regard to the implementation of the algorithm.
 In fact, the score function estimation problem
has been an important research topic in Bayesian statistics and machine learning \cite{hyvarinen2005estimation,vincent2011connection,alain2014regularized}.
In particular,  Alain and Bengio~\cite{alain2014regularized} showed that the minimization of the denoising autoencoder (DAE) objective function 
provides  an explicit way of approximating  the score function. This result was further
 extended using the amortized residual denoising autoencoder  (AR-DAE) for numerical stability and accuracy \cite{lim2020ar}.
 Interestingly, the training method for DAE or its variants is strikingly similar to that of Noise2X. Therefore,  by combining this with the Tweedie's formula, we can obtain a novel unified
 framework of Bayesian approach for self-supervised image denoising.
The conceptual difference and similarity of our method to the existing approaches are illustrated in Fig.~\ref{fig:concept}(a)(b). 
As the first step of our algorithm   is similar to Noise2X, we call our method Noise2Score.


%

Noise2Score is so powerful that it can be used to deal with any exponential family noises.
Moreover,  in contrast to the SURE approach, in which the network must be retrained using a different loss function depending on the noise model and parameters \cite{soltanayev2018training,kim2020unsupervised}, Noise2Score is universal in the sense that the identical neural network training step is used regardless of noise models.
 This property can be exploited to extend Noise2Score to blind setups in which noise parameters are unknown and should be estimated with minimal complexity.
In addition to the novel theoretical findings mentioned above, our empirical results using additive Gaussian, Poisson and Gamma noise models have shown that Noise2Score significantly outperforms SURE and Noise2X under similar experimental conditions.

%
%
%

\section{Related Works}

Here we give a brief overview of the existing deep learning approaches to denoise images.
As shown below, their main differences arise from the different choices of the distance metric $d(\cdot, \cdot)$
in Fig.~\ref{fig:concept}.

\subsection{Image denoising using supervised learning}

Let $\Xc$ and $\Yc$ denotes the spaces of clean and noisy images with the probability 
measure $P_X$ and $P_Y$, respectively.
When the paired ground-truth data $x$ is available for a noisy measurement $y$, i.e. $x:=x(y)$, then
 the supervised training can be carried out by minimizing the following loss function:
 \begin{eqnarray}
\ell_{sup}(\Theta) = \Ed_{y\sim P_Y}d(x, F_\Theta(y)),&\mbox{where} & d(x,F_\Theta(y)):=\|x - F_\Theta(y)\|^2 
\end{eqnarray}
where $F_\Theta(y)$ is a deep neural network parameterized by $\Theta$ with input $y$ and
here we use $l_2$ distance metric to simplify the explanation.
Although this approach has significantly improved performance compared to the classical denoising algorithms \cite{dabov2006image},
the supervised learning approach has a drawback as the large number of paired clean images are necessary. 
In real world applications, the acquisition for paired clean images are difficult or even impossible.

\subsection{Noise2X for image denoising without clean images}

To address this issue, various forms of the denoisers that can be trained without clean images have been proposed, which includes
Noise2Noise~\cite{lehtinen2018noise2noise}, Noise2Void~\cite{krull2019noise2void}, Noise2Self~\cite{batson2019noise2self}, Noise2Same ~\cite{xie2020noise2same}, etc.
These approaches are usually based on the variants of  loss functions that do not depend on clean images:
\begin{eqnarray}
\ell_{Noise2X}(\Theta) = \Ed_{y\sim P_Y}d(y', F_\Theta(y)),&\mbox{where} & d(y',F_\Theta(y)):=\|y' - F_\Theta(y)\|^2 
\end{eqnarray}
where the target image $y'$ is related to $y$ in unique ways depending on algorithms.
For example,  in Noise2Noise~\cite{lehtinen2018noise2noise}, $y'$ is 
 another noisy realization of the same underlying clean images. However,  Noise2Noise is not practical
 since multiple noisy realizations of the same image are not usually available in practice. 
 Noise2Void~\cite{batson2019noise2self} generates $y'$ by adding blind spots on $y$.
 In Noise2Self ~\cite{batson2019noise2self}, so-called J-invariant mask $J$  is added on the input $y$,
and $y'$ is defined as the image on the complementary mask. 
All of these changes of the target images are intended to prevent the network from converging to a trivial identity mapping.%
%
\subsection{Image denoising using Stein's Unbiased Risk Estimate (SURE)}

Rather than using the variants of a  loss function as in Noise2X,  
Soltanayev et al~\cite{soltanayev2018training} proposed a training scheme using the loss function from
Stein's unbiased risk estimate (SURE) \cite{stein1981estimation} 
which has additional regularization term.
Specifically, the loss function for the SURE denoiser is defined by 
\begin{eqnarray}\label{eq:SURE}
\ell_{SURE}(\Theta) =  \Ed_{y\sim P_Y} d(y, F_\Theta(y)),&~ d(y,F_\Theta(y)):= \|y- F_\Theta(y)\|^2+ 2\sigma^2 \mathrm{div}_y F_\Theta(y)  
\end{eqnarray}
Here,  the second term in $d(y,F_\Theta(y))$ of \eqref{eq:SURE} is the regularization term to prevent the network from converging to a trivial identity mapping,
where $\mathrm{div}_y$ denotes the divergence with respect to $y$.
%
Due to the difficulty of calculating the divergence term, the authors in \cite{soltanayev2018training} relied on MonteCarlo SURE \citep{ramani2008monte} which calculates the divergence term using MonteCarlo simulation.
This introduces additional hyperparameters, on which the final results critically depend.
Yet another limitation of SURE image denoiser is that the loss function is determined by the noise model.
For example, 
in \cite{kim2020unsupervised}, the authors derived Poisson Unbiased risk estimator (PURE) to train deep CNN for dealing with Poisson noise,
whose loss function is completely different from \eqref{eq:SURE}. Even under the same Gaussian noise models, if the noise variance $\sigma^2$ varies,
the neural network $F_\Theta$ must be trained again, which causes additional computational burden.
This differs from Noise2X, which uses the same loss function regardless of the noise models.

\section{Main Contribution: Noise2Score}

In contrast to the existing approaches shown in Fig.~\ref{fig:concept}(a),  our denoising approach  is composed of two steps  as illustrated in Fig.~\ref{fig:concept}(b),  which is inspired by the results from Bayesian statistics. 
In the following, we explain the details of each step.

\subsection{Tweedie's formula of the posterior mean for image denoising}

Suppose that the noisy measurement $y$ is given by
\begin{eqnarray}\label{eq:gaussian}
y=x+w,&\mbox{where}& w \sim \mathcal{N}(0,\sigma^2I)
\end{eqnarray}
where $x$ is the clean image, and $\mathcal{N}(0,\sigma^2I)$ denotes
the independent and identically distributed  (i.i.d) Gaussian distribution with zero mean and variance $\sigma^2$.
Then, Tweedie’s formula calculates the posterior expectation of $x$ given $y$ as \cite{robbins1956empirical}
\begin{eqnarray}\label{eq:post}
\Ed[x|y]= y+\sigma^2 l'(y),&\mbox{where}& l'(y)=\nabla_y \log p(y)
\end{eqnarray}
where $p(y)$ denotes the marginal distribution of $y$.
For the image corrupted by   Gaussian noises,
the posterior mean is the  minimum mean square error (MMSE) estimator, so that
 the denoised image can
be obtained  using \eqref{eq:post}, as long as
we know $l'(y)$ - the {\em score function} \cite{serfling2009approximation}.

Tweedie's formula was extended for general class of exponential family distribution \cite{efron2011tweedie}.
Specifically,
exponential family of probability distributions are defined as those
distributions whose density  have the following general form:
\begin{eqnarray}
p(y|\eta)=p_0(y)\exp\left(\eta^\top T(y) -\varphi(\eta) \right)
\end{eqnarray}
where the superscript $^\top$ denotes the transpose operation.
Here $\eta$ is a canonical (vector) parameter  of the family, $T(y)$ is a (vector) function of $y$,
 $\varphi(\eta)$ the cumulant generating
function which makes $p(y|\eta)$
 integrate to 1, and $p_0(y)$ the density up to a scale factor when $\eta = 0$. 
Exponential family distribution covers a large class of important distributions such as
the Gaussian, binomial, multinomial,
Poisson, gamma, and beta distributions, as well as many others.
Table~\ref{tbl:exp} summarizes the corresponding parameters for each exponential family distribution.

\begin{table*}[!b]
	\centering
	\caption{Tweedie's formula of exponential family distribution for image denoising. }
	\label{tbl:exp}
	\resizebox{\textwidth}{!}{
		\begin{tabular}{c|ccccc|c}
			\hline
			{Distribution}					& $p(y|x)$			&  $\eta$ & $T(y)$	 & $p_0(y)$   & $l_0'(y)$ & $\hat x$	\\ \hline\hline
			Gaussian  &  		$\frac{1}{\sqrt{2\pi}\sigma} e^{-\frac{(y-x)^2}{2\sigma^2}}$				& ${x}/{\sigma^2}$	& $y$	 &$\frac{1}{\sqrt{2\pi}\sigma} e^{-\frac{y^2}{2\sigma^2}}$	 	 & $-\frac{y}{\sigma^2}$  & $y+\sigma^2 l'(y)$ \\ 
			Gaussian  &  		$\frac{1}{\sqrt{2\pi}\sigma} e^{-\frac{(y-x)^2}{2\sigma^2}}$				& $\left[\frac{x}{\sigma^2}, -\frac{1}{2\sigma^2} \right]^\top$	& $\left[y,y^2\right]^\top$	 &$\frac{1}{\sqrt{2\pi}} $	 	 & $0$  & $y+\sigma^2 l'(y)$ \\ 
			Poisson	&	$\frac{x^ye^{-x}}{y!}$	& $\log(x)$	& $y$ 	&	  $\frac{1}{y!}$  & $\simeq -\log\left(y+\frac{1}{2}\right)$ &$\left(y+\frac{1}{2}\right)\exp(l'(y))$ \\ 
			Gamma($\alpha,\beta$) & $\frac{\beta^\alpha}{\Gamma(\alpha)}\left(\frac{y}{x}\right)^{\alpha-1}e^{-\beta\frac{y}{x}}$  & $\left[\alpha-1,-\frac{\beta}{x}\right]^\top$ & $\left[\log y,-y\right]^\top$ &  1  &0
			& $\frac{\beta y}{(\alpha-1)-y l'(y)}$ \\
			Bernoulli & $x^y(1-x)^{(1-y)}$ & $\log\left(\frac{x}{1-x}\right)$ &  $y$ & $1$ & 0 & $\frac{e^{l'(y)}}{1+e^{l'(y)}}$ \\
			Exponential & $xe^{-yx},~ y\geq 0$ &  $-x$ & $y$    & 1  & 0  & $-l'(y)$\\
			\hline
		\end{tabular}
	}
\end{table*}

Then, using the Bayes' rule, the posterior density of $\eta$ given $y$ is given by \cite{efron2011tweedie}:
\begin{eqnarray}
p(\eta|y) = \exp\left(\eta^\top T(y) -\lambda(y) \right)\left[p(\eta)e^{-\varphi(\eta)}\right],&\mbox{where}& \lambda(y)=\log\left(\frac{p(y)}{p_0(y)}\right)
\end{eqnarray}
where $p(y)$ and $p(\eta)$ denote the marginal distribution of $y$ and $\eta$, respectively.
This implies that the posterior density is again an exponential family distribution,
and the mode of the  posterior distribution  can be obtained by finding the maximum of $p(\eta|y)$.
Specifically, by computing the gradient of $\log p(\eta|y)$ with respect to $y$ and setting it to zero,  
the posterior estimate of the canonical parameter $\hat \eta$ should satisfy the following equality:
\begin{eqnarray}\label{eq:fix}
\hat\eta^\top  T'(y) &=&\lambda'(y)=-\nabla_y \log p_0(y)+\nabla_y \log p(y) 
=  -l_0'(y)+l'(y) 
\end{eqnarray}
where $l'(y):=\nabla_y \log p(y)$ and $l_0'(y):=\nabla_y \log p_0(y)$ are score functions, and $T'(y)=\nabla_y T(y)$.
In particular, if $T(y)=y$, we can obtain the following closed form solution for the posterior mean:
\begin{eqnarray}
\hat\eta:=\Ed[\eta|y]= \lambda'(y)
& =&  -l_0'(y)+l'(y)
\end{eqnarray}
In the following, we derive denoising algorithms using Tweedie's formulation  for several representative
distributions.

\paragraph{Additive Gaussian noise}

For  Gaussian distribution, in the first row of Table~\ref{tbl:exp}, we have 
\begin{eqnarray}
p_0(y)=\frac{1}{\sqrt{2\pi}\sigma} e^{-\frac{y^2}{2\sigma^2}}, && \eta = \frac{x}{\sigma^2}
\end{eqnarray}
Accordingly, 
\begin{eqnarray}
\Ed[\eta|y]= \frac{\hat x}{\sigma^2} =  \frac{y}{\sigma^2} + l'(y), &\mbox{where} & \hat x:= \Ed[x|y]
\end{eqnarray}
since $ l_0'(y)={y}/\sigma^2$. Therefore, the posterior mean  is given by
\begin{eqnarray}
\hat x = y + \sigma^2 l'(y)
\end{eqnarray}
which is equal to \eqref{eq:post}. Even with the different parameterization of Gaussian as shown
in the second row of Table~\ref{tbl:exp}, we can arrive at the same result using \eqref{eq:fix}.

\paragraph{Poisson noise}

For the case of Poisson noises, we have
\begin{eqnarray}
p_0(y)=\frac{1}{y!}&, &l_0'(y) =  -\frac{\nabla_y \Gamma(y+1)}{\Gamma(y+1)} \simeq -\log\left(y+\frac{1}{2}\right)  
\end{eqnarray}
where $\Gamma(y)$ denotes the gamma function and the last approximation comes from \cite{efron2011tweedie}.
Accordingly, we have
\begin{eqnarray}
\hat \eta = \log (\hat x)=  \log\left(y+\frac{1}{2}\right)  + l'(y)
\end{eqnarray}
This leads to  the following posterior estimate of the image
\begin{eqnarray}
\hat x = \left(y+\frac{1}{2}\right)\exp(l'(y))
\end{eqnarray}
This corresponds to Tweedie's formula for Poisson case~\cite{robbins1956empirical}.

This result can be generalized to the low photon count sensing scenario, where the sensor
measurement can be described by \cite{le2014unbiased}:
\begin{equation}
    y = \zeta z, \; z \sim \text{Poisson}(x/\zeta)
\label{poisson}
\end{equation}
where $z$ is a random variable that follows the Poisson distribution in Table~\ref{tbl:exp}
 and $\zeta >0$ is the gain of the acquisition process that is related to the noise level.
 Then, we have
 \begin{eqnarray}
p_0(z)=\frac{1}{z!}, && \eta = \log\left(\frac{x}{\zeta}\right)
\end{eqnarray}
Therefore, we have
\begin{eqnarray}
\frac{\hat x}{\zeta} =\left(z+\frac{1}{2}\right)\exp(l'(z)) & \longrightarrow &  {\hat x} =\left(y+\frac{\zeta}{2}\right)e^{ l'\left(\frac{y}{\zeta}\right)}
\end{eqnarray}

\paragraph{Gamma noise}

Gamma noise distribution can be used to model the speckle noises in various imaging application. 
Specifically, the image corrupted with the speckle noise is represented by
\begin{equation}
y = x n, \quad  n\sim p(n;\alpha,\beta) = \frac{\beta^\alpha}{\Gamma(\alpha)}n^{\alpha-1}\exp(-\beta n)
\end{equation}
where $p(n;\alpha,\beta)$ denotes the Gamma distribution with $(\alpha,\beta)$ parameters.
Here,  $(\alpha,\beta)$ are the parameters which determine the noise level of Gamma noise distribution.
For example, for the $k$-look measurement case, $\alpha=\beta=k$ \cite{bioucas2010multiplicative}.
Then, the probability $p(y|x)$ can be obtained by replacing $n$ in $p(n;\alpha,\beta)$ with
$n=y/x$, and the resulting density
function can be bound in Table~\ref{tbl:exp}.
Then, we can easily show that 
\begin{equation}
  p_o(y) = 1, \quad
 \eta = 
\begin{bmatrix}
\alpha-1 \\
-\beta/x
\end{bmatrix}, \quad
T(y) =
\begin{bmatrix}
\log y\\
y
\end{bmatrix}
\label{gamma}
\end{equation}
Using \eqref{eq:fix}, we have
\begin{eqnarray}\label{eq:gamma}
\frac{\alpha-1}{y} -\frac{\beta}{x} = l'(y)& \Longrightarrow & 
\hat x = \frac{\beta y}{(\alpha-1)-yl'(y)}
\end{eqnarray}

Similar derivations for other exponential family distribution
can be found in Table~\ref{tbl:exp}.

\subsection{Score function estimation}

So far, our derivation assumes the prior knowledge of the score function.
In  practice, this should be estimated, so here we describe how the score function can be estimated using a neural network. 

Historically, Hyv\"{a}rinen et al 
\cite{hyvarinen2005estimation}  was the first to derive a remarkable implicit
score matching objective  that no longer requires having an explicit score target  but is nevertheless equivalent to the original
problem that minimizes the expected quadratic distance between the model function and the score function of data. With the advance of the Denoising Auto Encoder (DAE) \cite{vincent2010stacked}, the author of \cite{vincent2011connection} observed that the minimization of the DAE objective function with the residual form is related to the score matching between the model and perturbed data.
This observation was rigorously analyzed by 
 Alain and Bengio~\cite{alain2014regularized}, who showed that 
as the perturbed noise becomes sufficiently small, DAE leads to the score function.

More specifically, in the denoising autoencoder (DAE), the following loss is minimized: 
\begin{equation}\label{eq:DAEcost}
    \ell_{DAE}(\Theta) = 
    \underset{\sigma_a \sim \mathcal{N}(0,\delta^2)}{\underset{u \sim \mathcal{N}(0,I)}{\underset{y \sim P_Y}{\Ed}}} \|y -F_\Theta(y+\sigma_a u)\|^2
\end{equation}
According to  \cite{alain2014regularized}, the optimal DAE $F_{\Theta^*}(x)$
can be represented by
\begin{equation}
F_{\Theta^*}(y) = y + \sigma_a^2 l'(y) + o(\sigma_a^2),   
\label{bengio}
\end{equation}
where $l'(y)$ is the score function defined in \eqref{eq:post}
and $o(\cdot)$ denotes the small ``o'' notation.
In other words, for sufficiently small $\sigma_a$, we can approximate the score function as:
\begin{eqnarray}\label{eq:DAEs}
\hat l'(y)= \frac{F_{\Theta^*}(y)-y}{\sigma_a^2}
\end{eqnarray}

One of the downsides of the score function estimation using DAE is that
the score function estimate in \eqref{eq:DAEs} is numerically unstable as $\sigma_a \rightarrow 0$.
To address the numerical instability and reduce the approximation error,
Lim et al \cite{lim2020ar} recently proposed so-called the amortized residual DAE (AR-DAE).
%
%
%
Specifically, AR-DAE is trained  by minimizing the following the objective function:
\begin{equation}
    \ell_{AR-DAE}(\Theta) = \underset{\sigma_a \sim \mathcal{N}(0,\delta^2)}{\underset{u \sim \mathcal{N}(0,I)}{\underset{y \sim P_Y}{\Ed}}}\|u + \sigma_a R_\Theta(y + \sigma_a u)\|^2
\label{eq:ardae}
\end{equation}
where  
$R_\Theta$ is  from the residual form of the DAE:
\begin{eqnarray}\label{eq:res}
F_\Theta(y)=\sigma_a^2 R_\Theta(y)+y
\end{eqnarray}
By plugging  \eqref{eq:res} in \eqref{eq:ardae}, we can obtain the original DAE cost in \eqref{eq:DAEcost} up to a scale factor.
Furthermore, using \eqref{eq:DAEs}, we have
\begin{eqnarray}\label{eq:gaus}
\hat l'(y)= \frac{F_{\Theta^*}(y)-y}{\sigma_a^2} = R_\Theta(y)
\end{eqnarray}
Therefore, 
the neural network trained with the AR-DAE
\eqref{eq:ardae} is a direct and stable way of estimating the score function.
Therefore, we employ the AR-DAE to estimate the score function from the noisy measurement as the first step of our method
(see Fig.~\ref{fig:concept}(b)).

\subsection{Relation to Noise2X and SURE}

Although our Noise2Score was derived from a completely different perspective, it turns out that it has very important connections to Noise2X and SURE.

Specifically, if the noisy image $y$ is corrupted by additive Gaussian,
using Tweedie's fomula for Gaussian noises in Table~\ref{tbl:exp}, the denoised image can be approximated by
\begin{eqnarray}\label{eq:xhat}
\hat x = y + \sigma^2 \hat l'(y) = F_{\Theta^*}(y)
\end{eqnarray}
where we use \eqref{eq:gaus}. 
Recall that $F_{\Theta^*}(y)$ is the neural nework trained with DAE loss function in \eqref{eq:DAEcost}.
The DAE training with \eqref{eq:DAEcost} is basically inserting noises into the images and then find the mapping that removes the noises. 
\add{In terms of adding extra noise to a noisy image, our method is closely related to Noisier2Noise~\cite{moran2020noisier2noise}. However, in the training phase of Noisier2Noise, the main assumption is that noise from the same noise statistic as the original noisy image should be injected into the noisy images so that the neural networks learn to reduce the noise from the noisy input image. In our Noise2Score training method, however, Gaussian noise is added to noisy images at different noise levels regardless of the noise statistics to estimate the score function rather than noise. In addition, in contrast to Noisier2Noise, our method has a post-processing step with the Tweedie’s formula, which is determined by the noise model such as Poisson, Gamma, etc.}

Yet another important connection is its relation to SURE.
Using the residual representation in \eqref{eq:res}, the 
SURE cost function in \eqref{eq:SURE} can be equivalently
represented by
\begin{eqnarray}
\ell_{SURE}(\Theta) &= &  \Ed_{y\sim P_Y} \left\{\|y- F_\Theta(y)\|^2+ 2\sigma^2 \mathrm{div}_y F_\Theta(y)  \right\} \notag \\
&= & \Ed_{y\sim P_Y} \left\{\sigma^4 \|R_\Theta(y)\|^2+ 2\sigma^4 \mathrm{div}_y R_\Theta(y)  \right\} +2\sigma^2 \mathrm{dim}(y)\
\end{eqnarray}
where $\mathrm{dim}(y)$ is the dimension of the vector $y$, which is constant.
Up to the scaling factor, this cost function is identical to the
implicit score matching objective by Hyv\"{a}rinen et al 
\cite{hyvarinen2005estimation}, which is given by
\begin{eqnarray}\label{eq:ISM}
\ell_{ISM}(\Theta) =  \Ed_{y\sim P_Y} \left\{ \frac{1}{2} \|\Psi_\Theta(y)\|^2+ \mathrm{div}_y \Psi_\Theta(y) \right\}
\end{eqnarray}
where $\Psi_\Theta(y)$ is the score function estimate parameterized by $\Theta$.
Therefore, using the residual transform in \eqref{eq:res}, the SURE objective is nothing but to find
the score function, which corresponds to the first step of our Noise2Score. Furthermore, due to the use of \eqref{eq:res}, SURE is not optimal for noise models other than Gaussian. \add{Although there are variants of SURE for other noise type~\cite{raphan2011least}, the cost function of this method varies depending on the type of noise. However, as discussed in detail later, our method uses the same cost function for neural network training, but only the post-processing step differs depending on the type of noise.} 
Moreover, considering that the modern research trend of the score function estimation has quickly evolved from the implicit score matching
to DAE or AR-DAE due to the numerical instability
and difficulty of computing the divergence term in \eqref{eq:ISM}, we can easily expect that
the SURE denoiser also suffers from similar limitations. 

\subsection{Universal neural estimation for blind denosing problems}

Recall that Noise2Score is composed of two steps: the score function estimation by minimizing the AR-DAE loss function,
after which the final denoised image is obtained by applying Tweedie's formula  shown in Table~\ref{tbl:exp}.
This decoupling allows Noise2Score to have unique advantages compared to the conventional  approaches.

Specifically, one of the most important advantages of Noise2Score is that
AR-DAE training is universal in the sense that
the same loss function is used regardless of noise models and parameters.
More specifically,  in  \eqref{eq:ardae},  the model
and noise parameter dependency is only through the sampling $y$ from
 $P_Y$. This sampling  step can be performed by random patch cropping from the input images, which procedure
 is identical regardless of $P_Y$ being Gaussian,  Poisson,  Gamma, etc.


Yet another advantage is that this property can be utilized to estimate the unknown noise parameters  without retraining neural networks.
For example, in the case of Gaussian denoising in Table~\ref{tbl:exp},
we have the  following form of the denoised images:
\begin{eqnarray}
\hat x(\sigma)= y +\sigma^2 \hat l'(y)
\end{eqnarray}
If the noise parameter $\sigma$ is unknown, then we can
find the optimal $\sigma^*$ value by solving the optimization problem
\begin{eqnarray}\label{eq:Q}
\sigma^*=\arg\min\limits_\sigma Q\left(\hat x(\sigma)\right)
\end{eqnarray}
where $Q(\cdot)$ is an image quality penalty such as total variation (TV).
In contrast to the SURE, where the $\sigma$ dependency exists in the loss function in \eqref{eq:SURE} and
the neural network must be trained with a new $\sigma$ value,
the optimization problem in \eqref{eq:Q} is  computationally negligible since the precalculated score function estimate $\hat l'(y)$ is 
used regardless of different $\sigma$ values.
A similar on-the-fly parameter estimation can be used for Gamma noise model. As shown in \eqref{eq:gamma}, the parameters
$(\alpha,\beta)$ can be adjusted using the precomputed score function estimate $\hat l'(y)$. Even for the Poisson noise model, the same trained neural network is used to compute the score function for each input $y/\zeta$.
\add{Therefore, our Noise2Score has advantages for the blind application where  existing methods  cannot be used due to the difficulty of  generating samples by changing the noise level parameters without retraining.}

\section{Experimental Results}
\paragraph{Dataset and Implementation detail}
We evaluated the proposed method for gray-scale and color images in the four benchmark datasets: the gray-scale image dataset contains BSD68~\cite{martin2001database} and Set12. The color image dataset contains RGB natural images CBSD68~\cite{martin2001database}, Kodak dataset. We adopt DIV2K~\cite{Timofte_2018_CVPR_Workshops}, and (C)BSD400 dataset as train data set. We generated the synthetic noise images for each noise distribution. In order to evaluate the proposed method fairly with comparison methods,  the same modified U-Net generator \cite{lehtinen2018noise2noise} is used for all methods. The total epoch was set to 100 and the Adam optimizer~\cite{kingma2014adam} was used to train the network. The learning rate was initialized to 2$\times10^{-4}$ for first 50 epoch, and after 50 epoch the learning rate was decayed to $2\times10^{-5}$. The proposed method was implemented in PyTorch~\cite{paszke2017automatic} with NVidia GeForce GTX 1080-Ti. The network training  took about 10 hours. \add{The more detail of implementation are described in Supplementary Material}. {To deal with blind noise case,  as for
the quality metric $Q(\cdot)$ we use the TV norm for the Gaussian noises and its combination with  data fidelity term for the Poisson and Gamma noise. The
details of the quality metric and the discussion on the accuracy of the blind parameter estimation is provided in  Supplementary Material.}

\begin{table}[t]
	\begin{small}
\begin{center}
			\caption{Quantitative comparison for various noise model using various methods in terms of PSNR(dB) when the noise parameters are known or unknown 
			(N2V: Noise2Void, N2S: Noise2Self, N2N: Noise2Noise, SL: supervised learning, Anscombe: Anscombe method for BM3D).}
				\label{tbl-results}
\resizebox{\textwidth}{!}
{%
\begin{tabular}{clcccccccccc}\toprule
	\multicolumn{1}{c}{Noise type}                    & & \multicolumn{7}{c}{Known parameters}                                     & \multicolumn{3}{c}{Unknown  parameters} \\ \cmidrule(r){3-9} \cmidrule(r){10-12}
	Gaussian                    & Dataset & BM3D     & N2V   & N2S   & SURE  & Ours           & N2N   & SL   & Ours        & N2N & SL   \\ \cmidrule(r){1-2} \cmidrule(r){3-9} \cmidrule(r){10-12}
	\multirow{4}{*}{$\sigma=$ 25}  & BSD68   & 28.59    & 26.77 & 28.28 & 28.78 & \textbf{29.12} & 29.18 & 29.20 & 28.95       & 28.98  & 28.98      \\
	& Set12   & 29.96    & 27.56 & 29.16 & 29.13 & \textbf{30.13} & 30.33 & 30.36 & 30.08       & 30.08   & 30.08    \\
	& CBSD68  & 30.56    & 29.22 & 30.05 & 30.23 & \textbf{30.85} & 31.10 & 31.10 & 30.78       & 30.91   & 30.91    \\
	& Kodak   & 31.68    & 30.02 & 30.53 & 30.75 & \textbf{31.89} & 32.20 & 32.20 & 31.78       & 31.96   & 31.96     \\ \cmidrule(r){1-2} \cmidrule(r){3-9} \cmidrule(r){10-12}
	\multirow{4}{*}{$\sigma$ = 50}  & BSD68   & 25.62    & 24.34 & 25.61 & 25.80 & \textbf{26.21} & 26.27 & 26.30 & 25.81       & 25.86  & 25.98     \\
	& Set12   & 26.33    & 24.68 & 26.19 & 26.23 & \textbf{27.16} & 27.20 & 27.20 & 26.65       & 26.65  & 26.72    \\
	& CBSD68  & 27.38    & 25.13 & 27.05 & 26.24 & \textbf{27.75} & 27.94 & 27.95 & 27.32       & 27.66  & 27.68     \\
	& Kodak   & 27.02    & 25.75 & 28.01 & 26.93 & \textbf{28.83} & 29.07 & 29.10 & 28.13       & 28.70  & 28.71     \\ \cmidrule(r){1-2} \cmidrule(r){3-9} \cmidrule(r){10-12}
	Poisson                    & Dataset & Anscombe & N2V   & N2S   & PURE  & Ours           & N2N   & SL   & Ours  & N2N   & SL         \\ \cmidrule(r){1-2} \cmidrule(r){3-9} \cmidrule(r){10-12}
	\multirow{4}{*}{$\zeta$ = 0.01} & BSD68   & 30.51    & 28.73 & 29.76 & 30.16 & \textbf{30.81} & 30.89 & 30.98 & 30.63       & 30.91  & 30.93     \\
	& Set12   & 31.20    & 30.06 & 30.47 & 30.69 & \textbf{31.58} & 31.70 & 31.79 & 31.42       & 31.67  & 31.70     \\
	& CBSD68  & 32.40    & 31.85 & 31.04 & 32.30 & \textbf{32.61} & 33.01 & 33.01 & 32.23       & 32.94  & 32.95     \\
	& Kodak   & 33.13    & 32.98 & 32.24 & 33.01 & \textbf{33.41} & 33.91 & 33.91 & 32.96       & 33.85  & 33.87     \\ \cmidrule(r){1-2} \cmidrule(r){3-9} \cmidrule(r){10-12}
	\multirow{4}{*}{$\zeta$ = 0.05} & BSD68   & 26.77    & 26.12 & 26.54 & 24.76 & \textbf{27.12} & 27.23 & 27.25 & 26.77       & 27.19  & 27.20      \\
	& Set12   & 27.54    & 27.21 & 27.53 & 25.07 & \textbf{27.86} & 28.02 & 28.04 & 27.76       & 28.01  & 28.03     \\
	& CBSD68  & 28.33    & 28.37 & 28.32 & 26.68 & \textbf{28.68} & 29.23 & 29.27 & 28.23       & 29.07  & 29.07     \\
	& Kodak   & 29.31    & 29.56 & 29.52 & 26.38 & \textbf{29.71} & 30.33 & 30.40 & 28.98       & 30.13  & 30.25     \\ \cmidrule(r){1-2} \cmidrule(r){3-9} \cmidrule(r){10-12}
	Gamma                   & Dataset &       & N2V   & N2S   &     & Ours           & N2N   & SL   & Ours  & N2N  & SL         \\ \cmidrule(r){1-2} \cmidrule(r){3-9} \cmidrule(r){10-12}
	\multirow{4}{*}{$k$ = 100}    & BSD68   & -        & 29.32 & 30.49 & -     & \textbf{32.67} & 32.87 & 32.93 & 32.54       & 32.83 & 32.88      \\
	& Set12   & -        & 30.54 & 30.71 & -     & \textbf{33.01} & 33.21 & 31.72 & 32.89       & 33.19 & 33.21       \\
	& CBSD68  & -        & 31.11 & 30.54 & -     & \textbf{33.82} & 35.45 & 35.53 & 33.50       & 35.33  & 35.33     \\
	& Kodak   & -        & 31.96 & 31.60 & -     & \textbf{34.22} & 36.26 & 36.41 & 33.82       & 36.16   & 36.16    \\ \cmidrule(r){1-2} \cmidrule(r){3-9} \cmidrule(r){10-12}
	\multirow{4}{*}{$k$ = 50}     & BSD68   & -        & 26.98 & 29.25 & -     & \textbf{30.53} & 31.10 & 31.16 & 30.42       & 30.64 & 30.64      \\
	& Set12   & -        & 27.36 & 29.67 & -     & \textbf{30.87} & 31.68 & 31.72 & 30.83       & 31.58  & 31.67     \\
	& CBSD68  & -        & 30.51 & 30.19 & -     & \textbf{31.05} & 33.52 & 33.53 & 30.93       & 33.40  & 33.41     \\
	& Kodak   & -        & 31.38 & 31.03 & -     & \textbf{31.34} & 34.49 & 34.57 & 31.32       & 34.39  & 34.40   \\ \bottomrule
\end{tabular}}
	\end{center}
	\end{small}
\end{table}

\begin{figure}
    \centering
	\includegraphics[width=0.9\linewidth]{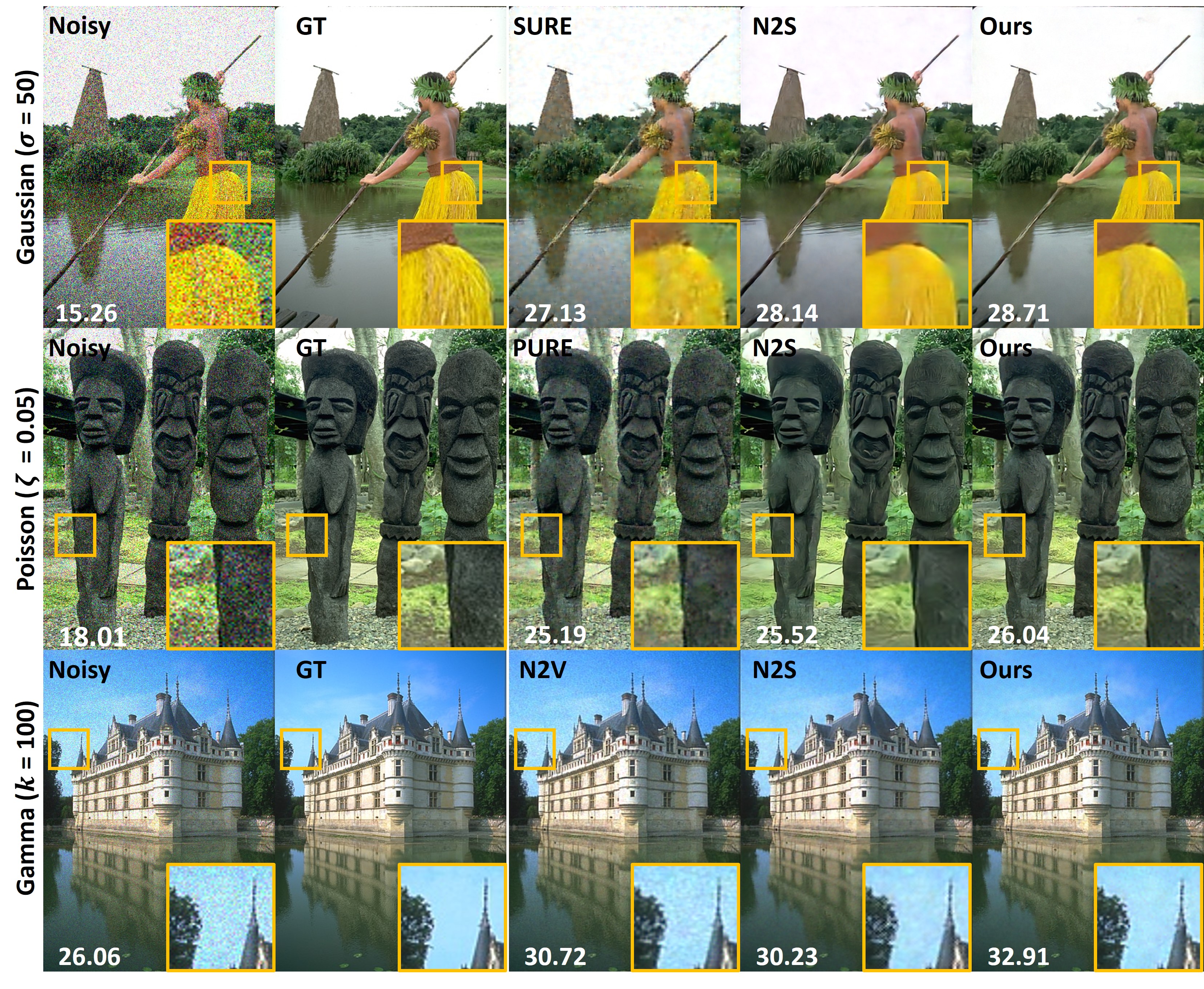}
    \caption{Qualitative comparison using CBSD68 dataset. Top : Gaussian noise  with $\sigma$ = 50. Middle: Poisson noise with $\zeta$ = 0.05. Bottom: Gamma noise with $k$ = 100. White numbers at the lower left part of the images indicate the PSNR values in dB.  Noisy: noisy input,  GT:  ground-truth image.}
        \label{fig:result}
\end{figure}

\paragraph{Gaussian noise}
Using additive Gaussian noise, our method was compared to (C)BM3D \cite{dabov2006image}, Noise2Void, Noise2Self, SURE, Noise2Noise and supervised  learning  as shown in Table 2. 
As expected, the supervised learning and  Noise2Noise using multiple noisy realization were the best, but these methods are not practical as explained before.
Among the other self-supervised learning approaches, our Noise2Score provide the best performance for all dataset.
With the comparison with the supervised learning and Noise2Noise, the performance difference of our method is only margin.
Even for blind-noise cases where the noise levels are randomly sampled from Gaussian distribution with $\sigma \in$ [5, 55],
Noise2Score with the on-the-fly parameter estimation
performed comparable to the supervised learning approach. These results indicate the superior performance of our method.     
The qualitative comparison in Fig.~\ref{fig:result} show that Noise2Score retains the better image details and provides
much visually pleasing results compared to other top two self-supervised learning approaches.

\paragraph{Poisson noise}
For the Poisson noise case, BM3D and SURE were replaced with Anscombe (BM3D+VST)~\cite{makitalo2010optimal} and  PURE \cite{kim2020unsupervised}, respectively,  as indicated in Table 2, as  they are optimized for Possion noise model.
Again, our Noise2Score results in significant performance gain compared to other approaches,
and the results are comparable to supervised learning approach and Noise2Noise.
In addition, to perform experiments when the noise parameters are unknown, we performed experiments with  the noise levels randomly sampled between $\zeta \in$ [0.001,0.1]. We found that our proposed method produce even higher PSNR than the results of  other existing methods even with known parameters,
and also provide comparable results to the supervised learning approaches. 
The qualitative comparison in Fig.~\ref{fig:result} with the other two best approach showed that
our proposed method provides much cleaner images.

\paragraph{Gamma noise}
As the extension of BM3D and SURE are not available for Gamma noises,
only four comparison methods was adopted for comparison as indicated in Table~\ref{tbl-results}.  Here, we set $\alpha=\beta=k$.
Again, 
 Noise2Score produced best results among self-supervised learning approaches, and provided  comparable results to  Noise2Noise and supervised learning approaches.
Even for the blind noise case, where the noise parameter $k$ randomly selected in $k \in [40,120]$,  Noise2Score provided the better PSNR results
compared to the other self-supervised learning method even with known parameters, and produces comparable results to supervised learning approaches.
 The qualitative comparison in Fig.~\ref{fig:result} confirm that our Noise2Score provides best reconstruction results.

\section{Conclusion}

In this work, we  provided a novel Bayesian framework for self-supervised im\label{key}age denoising without clean data, which surpasses SURE, PURE, Noise2X, etc.  Our novel innovation came from  the Tweedie's formula,
which provides explicit representation of denoise images through the score function. By combining with the score-function estimation using AR-DAE, our Noise2Score
can be applied to image denoising problem from any exponential family noises. Furthermore,  an identical neural network training can be
universally used regardless of the noise models, which leads to  the noise parameter estimation with minimal complexity.
The links to SURE and existing Noise2X were also explained, which clearly showed why our method is a better generalization.

\section*{Limitation and negative societal impacts} 
As a negative societal impact, the failure of image denoising methods could produce side effects. For example, removing both the noise and the texture of the medical images could lead to misdiagnoisis. While the proposed Noise2Score has merits, there are also limitations. In real environments, the prior knowledge of noise distribution may not be available, and the noise model could not be modeled by exponential family noises. Therefore, future extension to such scenario would be beneficial.

\section*{Acknowledgment} 
This research was funded by the National Research Foundation (NRF) of Korea grant NRF-2020R1A2B5B03001980,  ETRI (Electronics and Telecommunications Research Institute)’s internal funds [21YR2500, Development of Digital Biopsy Core Technology for high-precision Diagnosis and Therapy of Senile Disease], and KAIST Key Research Institute (Interdisciplinary Research Group) Project.
This work was also supported by Institute of Information $\&$ Communications Technology Planning $\&$ Evaluation (IITP) grant funded by the Korea government (MSIT) (No.2019-0-00075, Artificial Intelligence Graduate School Program (KAIST)).

{\small
\bibliographystyle{unsrt}

}

\appendix
\section*{Appendix}
\section{Algorithm of Noise2Score}
Algorithm~\ref{algoritm} details the overall pipeline of the Noise2Score. First, the neural network $R_{\Theta}$ was trained by minimizing $\ell_{AR-DAE}(\Theta)$ to learn the estimation of the score function from the noisy input $y$.
This neural network training step is universally applied regardless of noise distribution $\eta$. In particular,
during the training phase, we annealed $\sigma_a$ from $\sigma_a^{max}$ to $\sigma_a^{min}$ to stably train the network as suggested in~\cite{lim2020ar}. After training the network $R_{\Theta}$, we estimated the clean images for each noise distribution by using Tweedie's formula as reported in Table 1 in the main paper. 

\begin{algorithm}
	\caption{Noise2Score}
	\label{algo-1}
	\SetKwInOut{Input}{Input}
	\SetKwInOut{kwtest}{Inference}
	\SetKwInput{kwInit}{Output}
	\SetKwInput{kwset}{Given}
	\kwset{learning rates  $\rho$, number of epochs $N$;}
	\Input{noisy input $y$ from training data set $D_{\eta}$ with size $m$  and noise level parameter $\eta$~$\in (\sigma, \zeta, k$), neural network  $R_{\Theta}$, annealing sigma $\sigma_a \in [\sigma_a^{min},\sigma_a^{max}]$;}
	\For {$n=1$ to N}
	{
		$u \sim \mathcal{N}(0,1)$;\\
		$q \rightarrow n/m$; \\
		$\sigma_a \rightarrow \sigma_a^{max}*(1-q) + \sigma_a^{min}*q$ \\
		$\ell_{AR-DAE}(\Theta) ={\underset{u \sim \mathcal{N}(0,I),\sigma_a \sim \mathcal{N}(0,\delta^2)}{\underset{y \sim P_Y}{\Ed}}}\|u + \sigma_a R_\Theta(y + \sigma_a u)\|^2$;\\
		$\Theta \gets \Theta - \rho \nabla_{\Theta} \ell_{AR-DAE}(\Theta)$;\\
		
	}
	\kwInit{estimated the score function, $ R_{\Theta}(y)$ = $\hat l'(y)$ }
	\kwtest{}
	\lIf {\mbox{\rm Gaussian noise}}{$\hat{x}$ = $ y + \sigma^2 l'(y)$}
	\lElseIf {\mbox{\rm Poisson noise}}{$\hat{x}$ =$\left(y+\frac{\zeta}{2}\right)e^{ l'\left(\frac{y}{\zeta}\right)}$}
	\lElseIf {\mbox{\rm Gamma noise} }{$\hat{x}$ = $\frac{\beta y}{(\alpha-1)-yl'(y)}$}	
	\label{algoritm}
\end{algorithm}

\begin{figure}[h]
	\vspace{-1em}
	\centering	
	\begin{subfigure}{.32\linewidth}
		\centering
		\includegraphics[width=1\linewidth]{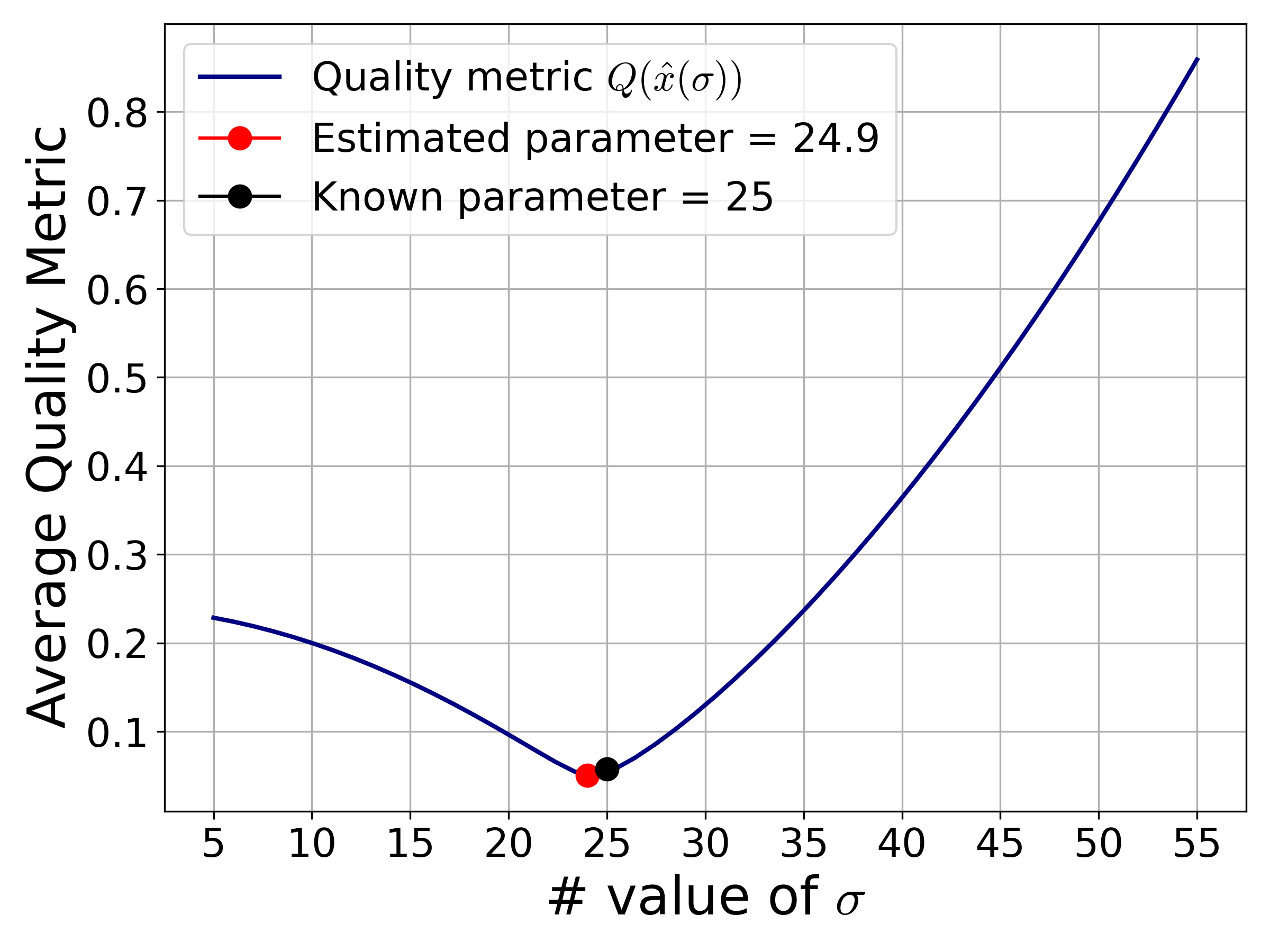}  
		\caption{Gaussian}
		\label{fig:gau}
	\end{subfigure}
	\begin{subfigure}{.32\linewidth}
		\centering
		\includegraphics[width=1\linewidth]{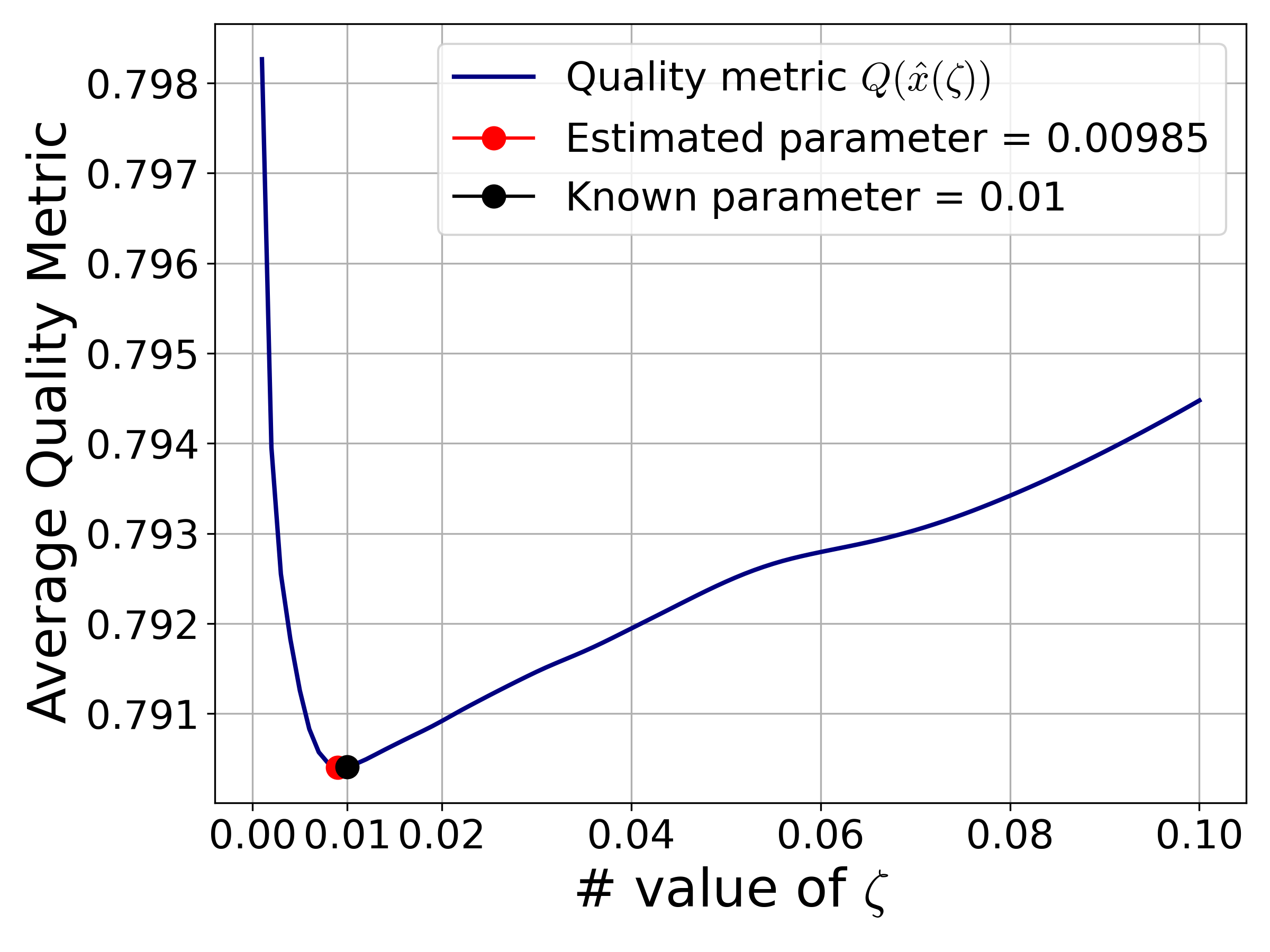}  
		\caption{Poisson}
		\label{fig:poi}
	\end{subfigure}
	\begin{subfigure}{.32\linewidth}
		\centering
		\includegraphics[width=1\linewidth]{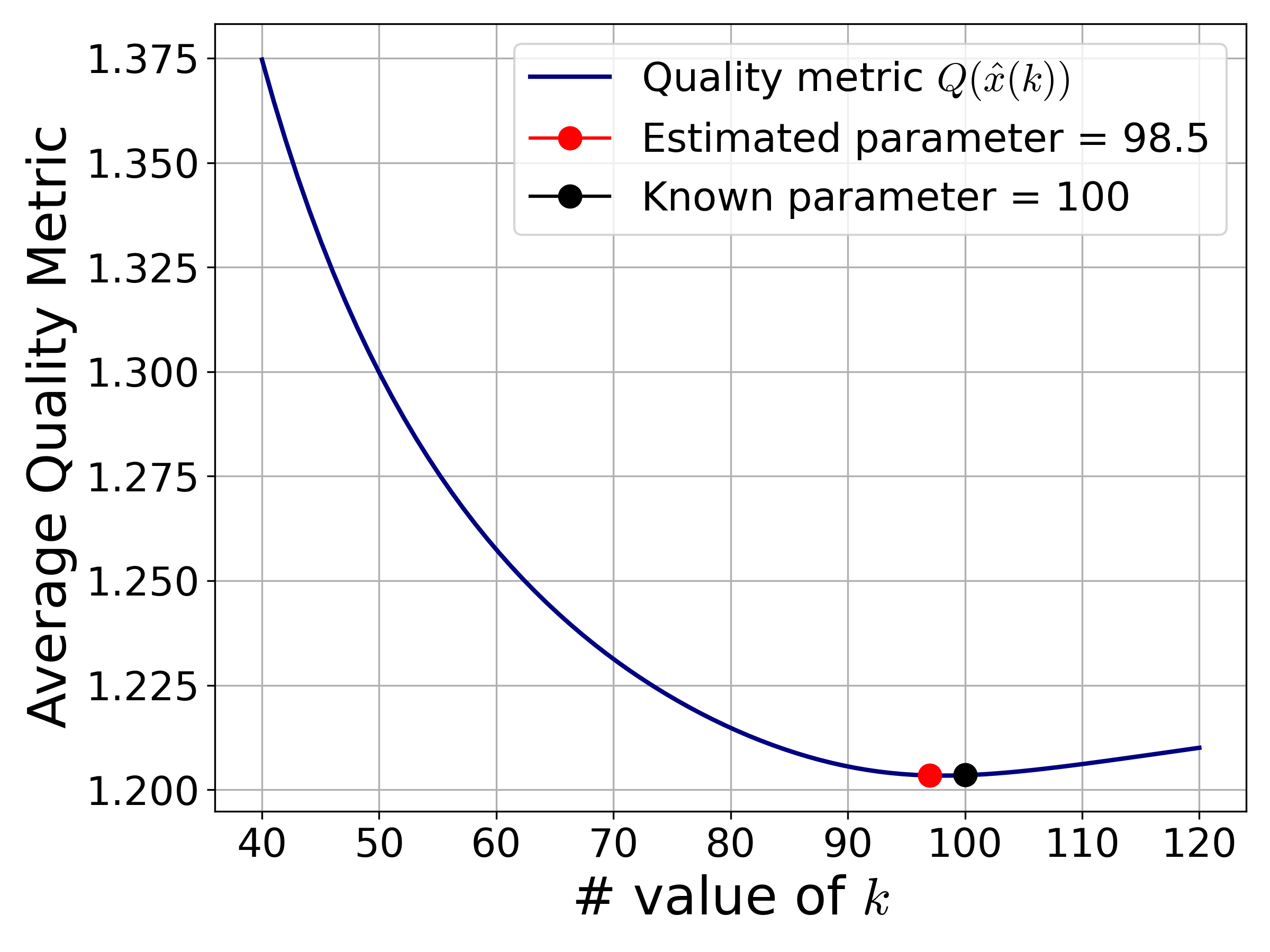}  
		\caption{Gamma}
		\label{fig:gamma}
	\end{subfigure}	
	\caption{The average quality penalty value with respect to the noise level parameter such as $\sigma, \zeta, k$ in the Set12 dataset. (a)  Gaussian noise with $\sigma = 25$, (b) Poisson noise with $\zeta = 0.01$, (c) Gamma noise with $\alpha=\beta=k$ = 100. The red dot indicate the estimated noise level parameter by our method, whereas the black dots indicate the ground-truth parameters.}
	\label{fig:estimation}
\end{figure}

\section{Quality metric for blind noise removal}
To deal with blind noise removal case where the noise parameters are unknown, the image quality penalty $Q(\cdot)$ should be defined for each noise distribution as stated in the main paper. Quality metrics was built upon the total variation norm, but  
the data fidelity term of noise distribution was combined with total variation in the case of Poisson and Gamma noise as suggested in
\cite{chan2007multilevel,chen2019convex}. The details are as follows.

\textbf{Additive Gaussian noise}
For the addtivie Gaussian noise case, we  design the quality metric by using just the total variation norm: 
\begin{eqnarray}
	Q(\hat x(\sigma)) = |\nabla \hat{x}(\sigma)|    
\end{eqnarray}
Fig.~\ref{fig:gau} shows the results of the estimation of an unknown parameter for Gaussian noise $\sigma$ = 25. Esitmated parameter $\sigma^{\ast}$ from quality metric for Gaussian noise, $Q(\hat x(\sigma))$, is 24.8 compared to target parameter $\sigma = 25$, which confirms that our blind
noise parameter estimation is quite accurate.

\textbf{Poisson noise}
In case of the Poisson noise, we employ the existing total variation regularization method that jointly minimizes the negative log likelihood of the prior distribution from the EM algortihm~\cite{chan2007multilevel}. Specifically,
the quality metric for Poisson noise $Q(\hat x(\zeta))$ was defined as follow:

\begin{equation}
	Q(\hat x(\zeta)) = \alpha |\nabla\hat{x}(\zeta)| + \hat{x}(\zeta) - y\log\hat{x}(\zeta) 
\end{equation}
where $\alpha$ is the hyperparameter for the weighting the total variation norm. Following the paper~\cite{chan2007multilevel}, the $\alpha$ value was set to 0.25, 0.1 in the case of $\zeta = 0.05, 0.01$, respectively.
Fig.~\ref{fig:poi} shows the results of the estimation of an unknown parameter for Poisson noise $\zeta$= 0.01. Estimated parameter $\zeta^{\ast}$ from quality metric for Poisson noise, $Q(\hat x(\zeta))$, is 0.00985 compared to target parameter $\zeta = 0.01$. This again confirms
the accuracy of our blind parameter estimation.

\textbf{Gamma noise}
For the Gamma noise, we employ the method  from the MAP estimation of Gamma noise distribution~\cite{chen2019convex}. 
Specifically, the quality metric for Gamma nose with $\alpha=\beta=k$,  $Q(\hat x(k))$, was defined as follow: 
\begin{equation}
	Q(\hat x(k)) = |\nabla\hat{x}(k)| +\alpha\frac{y}{\hat{x}(k)} + \frac{\beta}{2}\left(\frac{y}{\hat{x}(k)}\right)^2 +\gamma \log \hat{x}(k)
\end{equation}
where $\alpha, \beta, \gamma$ are the hyperparameter for weighting each term. In  \cite{chen2019convex}, the authors reported that
when $\alpha + \beta = \gamma$ the results are stable, so that we follow this condition and each parameter was set to 0.5, 0.5, and 1, respectively.  
Fig.~\ref{fig:gamma} shows the results of the estimation of an unknown parameter for Gamma noise $k$= 100. Estimated parameter $k^{\ast}$ from quality metric for Gamman, $Q(\hat x(k))$, is 98.5 compared to target parameter $k = 100$. From  Fig~\ref{fig:estimation}c, we found that our proposed quality metric for each noise distribution was quite good to find the optimal parameter to get the best performance in case of the blind noise.

\section{Detail of Dataset}
We adopted  DIV2K and CBSD400 dataset as the trainset. We generated the cropped patch 128$\times$128 size to train the network. For the data augmentation, we used the random horizontal, vertical filp and flop methods. To evaulate the proposed Noise2Score, the testset was adopted for Kodak, CBSD68, Set12. In case of the BSD68 dataset, we transform the CBSD68 datset into gray-scale images. The URL for each dataset is given by:

DIV2K :~  \href{https://data.vision.ee.ethz.ch/cvl/DIV2K/}{https://data.vision.ee.ethz.ch/cvl/DIV2K/} 

CBSD400 :~
\href{https://www2.eecs.berkeley.edu/Research/Projects/CS/vision/bsds/}{https://www2.eecs.berkeley.edu/Research/Projects/CS/vision/bsds/}

Kodak :~ \href{http://www.cs.albany.edu/~xypan/research/snr/Kodak.html/}{http://www.cs.albany.edu/~xypan/research/snr/Kodak.html/} 

Set12 :~
\href{https://www.researchgate.net/figure/12-images-from-Set12-dataset_fig11_338424598}{https://www.researchgate.net/figure/12-images-from-Set12-dataset/fig11/338424598}

\section{Implementation Detail}
\add{
In the training phase, we only selected one noise distribution. For example, if we train Noise2Score for Gaussian noise, all training images are corrupted by Gaussian noise. In the case of the "known" parameter, during the training we sampled noisy images with only one noise level, but in the case of experiments with unknown noise parameters, we randomly sampled images with multiple noise levels during training, as reported in the main paper. In orther wordks, the multiple trained U-Net are required to deal with the multiple cases of the combined the noise model and the noise level. For the annealing sigma $\sigma_a$ which are need to learn the score function of noisy data, we linearly decrease the perturbed noise from $\sigma_a^{max}$ to $\sigma_a^{min}$. In the cases of the Gaussian and Gamma noise, $\sigma_a^{max}$ and $\sigma_a^{min}$ are set to [0.1, 0.001], respectively. Due to the normalized issue of intensity range, we set $\sigma_a^{max}$ and $\sigma_a^{min}$ to [1, 0.05] for the Poisson noise case.   
}
\section{Comparison with More Baseline Algorithms}
\add{
To show the effectiveness of proposed Noise2Score, we additionally carried out experimental comparisons with Laine et al~\cite{laine2019high}, Noisier2Noise~\cite{moran2020noisier2noise} and Noise2Same\cite{xie2020noise2same}. For a fair comparison, we trained all methods with the combined DIV2K and BSD400 dataset in gray scale. Then, the trained networks were used for inference using the BSD68 dataset for Gaussian noise and Poisson noise, as shown Table 3. The results confirmed that our method outperformed the existing ones including two different implementations of Laine et al, Noisier2Noise and Noise2Same. 
}
\begin{table}[h]
\begin{center}
\begin{small}
\caption{Quantitative comparison for various noise model using various methods in terms of PSNR(dB) when the noise parameters are known (N2V: Noise2Void, N2S: Noise2Self, Nr2N: Noisier2Noise, N2Same: Noise2Same).}
\begin{tabular}{cccccccc} \toprule
Noise type & N2V & N2S & Nr2N & N2Same & Laine19-mu & Laine19-pme & Ours  \\ \cmidrule(lr){1-1} \cmidrule(lr){2-7} \cmidrule(lr){8-8}
Gaussian ($\sigma$ = 25) & 26.27 & 28.28      & 28.01         & 28.00      & 28.13      & 29.04   & \textbf{29.12} \\
\cmidrule(lr){1-1} \cmidrule(lr){2-7} \cmidrule(lr){8-8}
Poisson ($\zeta$= 0.01) & 28.73  & 29.73      & -           & 29.32      & 28.49      & 30.70   & \textbf{30.81} \\ \bottomrule
\end{tabular}
\end{small}
\end{center}
\label{table:more}
\end{table}

\section{Application on Real Data}
\add{
Although the proposed Noise2Score has difficulties in application in real data sets, we have shown the potential of Noise2Score for real application.
We carried out the experiments with real fluorescence microscopy data sets (FMD) as shown Table 4. We have taken raw noisy images from the confocal FISH categories, which consisted of 1000 images. We used the 900 images as a training data set and the remaining 100 images as a test data set. In the FMD data set, the noisy images are modeled with mixed Poisson-Gaussian noise. Accordingly, we used a two-step approach in which the Gaussian noise is first removed using the Tweedie’s formula for the Gaussian case and the Poisson noise is subsequentially reduced using the Tweedie's formula for the Poisson noise. As the same trained neural network is used for both steps, the computational complexity increase is negligible. The results in Table 4 indicated that our method still outperformed other methods.
}
\begin{table}[h]
\begin{center}
\begin{small}
\caption{Comparison results in FMD data set with real noises in terms of PSNR (dB).}
\begin{tabular}{ccccc} \toprule
Dataset & N2V & N2S & N2Same & Ours  \\ \cmidrule(lr){1-1} \cmidrule(lr){2-4} \cmidrule(lr){5-5}
Confocal MICE & 35.83 & 36.32 & 36.42  & \textbf{36.73} \\
\bottomrule
\end{tabular}
\end{small}
\end{center}
\label{table:mICE}
\end{table}
\
\section{Case Study for Noise Statics and Noise Level Mismatch}
\add{
In this section, we examined the mismatch of the noise statisticcs and the noise level between the training data and the inference level. 
For the mismatch of noise model, we carried out experiments in which the noise statistics of the training data differ from those in the inference phase. As shown in Table 5, all denoisers perform best when the noise statistics match in the training and inference phases. If the noise statistics differ between the two phases, we can observe  performance deterioration in all methods. Nevertheless, the proposed method surpasses the other methods in almost the cases.
}

\add{
In the case of the mismatch of noise level, we additionally carried out experiments when there is a noise level mismatch between the training and inference phases. Table 6 shows comparison results using BSD68 dataset. Specifically, all methods are trained with a training data set corrupted with Poisson noise distribution with $\zeta$ = 0.01. If the image at the inference phase is also corrupted with the Poisson noise with $\zeta$ = 0.01, all methods show the best performance. However, as the noise level in the test data increases, their performance decreased. Nonetheless, our proposed method shows the most robust performance compared with other self-supervised methods.
}
\begin{table}[h]
\begin{center}
\begin{small}
\caption{Comparison with different methods in the event of a discrepancy between the noise statistics of the training set and the test set in the CBSD68 data set.}
\begin{tabular}{cccc} \toprule
Test set of noisy type & Gaussian ($\sigma$ = 25) & Poisson ($\zeta$ = 0.01 ) & Gamma ($k$ = 100) \\ \cmidrule(lr){1-1} \cmidrule(lr){2-4} 
Train set of noisy type  & N2V / N2S / Ours & N2V / N2S / Ours & N2V / N2S / Ours  \\ \cmidrule(lr){1-1} \cmidrule(lr){2-4}
Gaussian ($\sigma$ = 25)  & 29.22 / 30.05 / 30.85 & 29.69 / 30.23 / 30.65 & 30.32 / 29.74 / 29.57  \\
Poisson ($\zeta$ = 0.01 ) & 26.73 / 26.43 / 27.65  & 31.85 / 31.04 / 32.61 & 30.21 / 30.63 / 31.23 \\
Gamma ($k$ = 100) & 25.74 / 26.51 / 27.95 &	29.06 / 30.34 / 29.34 &	31.14 / 30.54 / 33.82 \\
\bottomrule 
\end{tabular}
\end{small}
\end{center}
\label{table:mICE}
\end{table}

\begin{table}[h]
\begin{center}
\begin{small}
\caption{Quantitative results with various methods when there is a noise level mismatch between the training data set and test data in the BSD68 dataset.}
\begin{tabular}{cccccc} \toprule
&\multicolumn{5}{c}{Inference noise level $\zeta$}  \\ \cmidrule(lr){2-6} 
Method & 0.01 & 0.02 & 0.03 & 0.04 & 0.05  \\ \cmidrule(lr){1-1} \cmidrule(lr){2-6} 
N2V & 28.73 & 26.67 & 24.79  & 23.39 & 22.32 \\
N2S & 29.76 & 27.03 & 24.73  & 23.01 & 21.71\\
Ours & \textbf{30.81} & \textbf{29.01} & \textbf{28.00}  & \textbf{27.25} & \textbf{26.64}\\
\bottomrule
\end{tabular}
\end{small}
\end{center}
\label{table:mICE}
\end{table}

\section{Qualitative Results}
\begin{figure}[!b]
	\centering
	\includegraphics[width=1\linewidth]{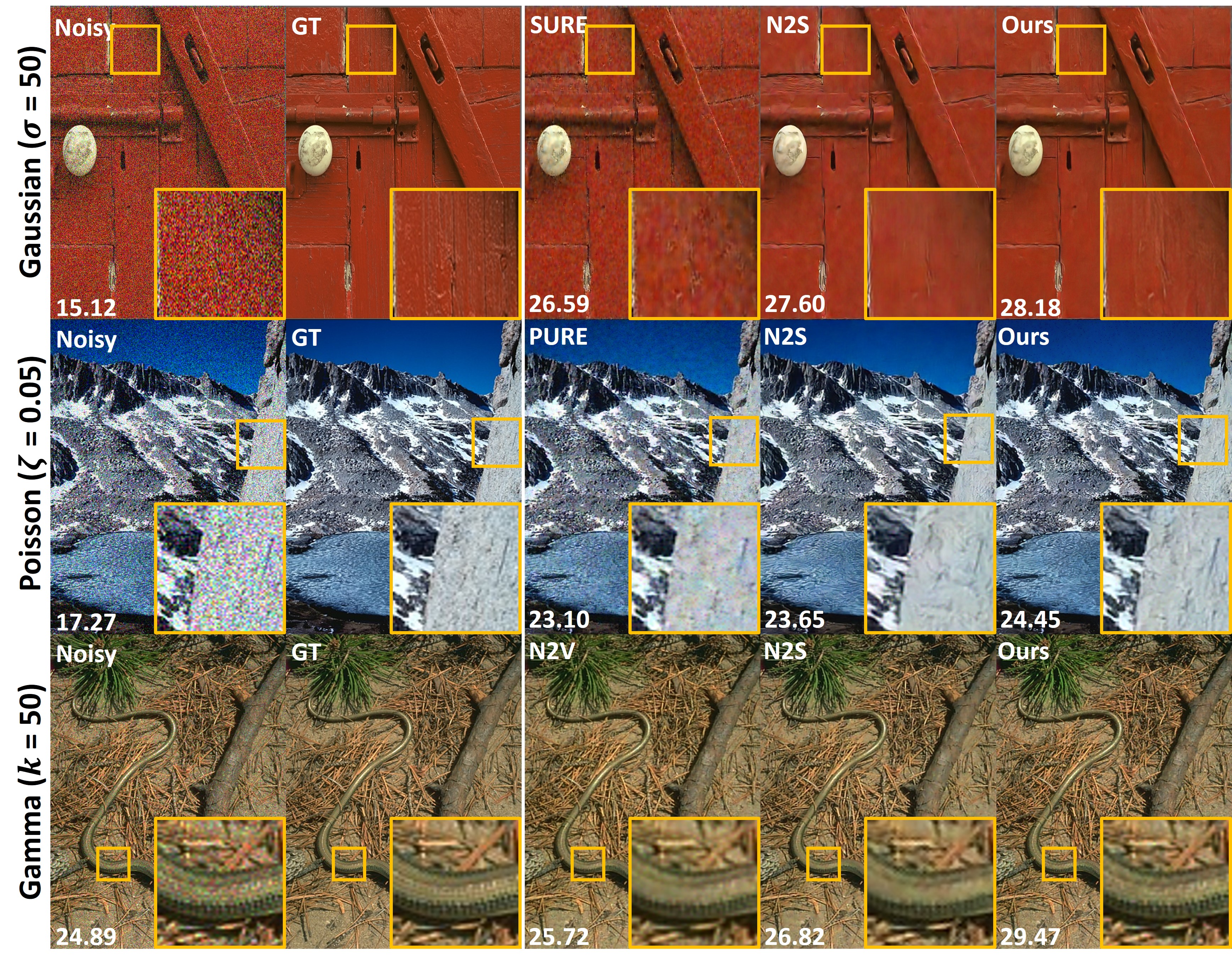}  
	\caption{Qualitative comparison using Kodak and CBSD68 dataset. Top : Gaussian noise with $\sigma$= 50. Middle: Poisson noise with $\zeta$ = 0.05. Bottom: Gamma noise with $k$ = 50. White numbers at the lower left part of the images indicate the PSNR values in dB. Noisy: noisy input, SURE: Stein unbiased risk estimate based denoiser, PURE: Poisson SURE, N2V: Noise2Void, N2S: Noise2Self.}
	\label{fig:non}
\end{figure}

We provided an additional qualitative comparison to valid the effectiveness of the proposed method. Fig.~\ref{fig:non} illustrates the visual results of image denoising with various methods on the color image dataset, such as Kodak and CBSD68 dataset, when the noise parameters
are known. Compared to the SOTA self-supervised learning approaches, our method provides 0.5$\sim$3dB gain in PSNR
with superb subjective image qualities.

\begin{figure}[!t]
	\centering
	\includegraphics[width=1\linewidth]{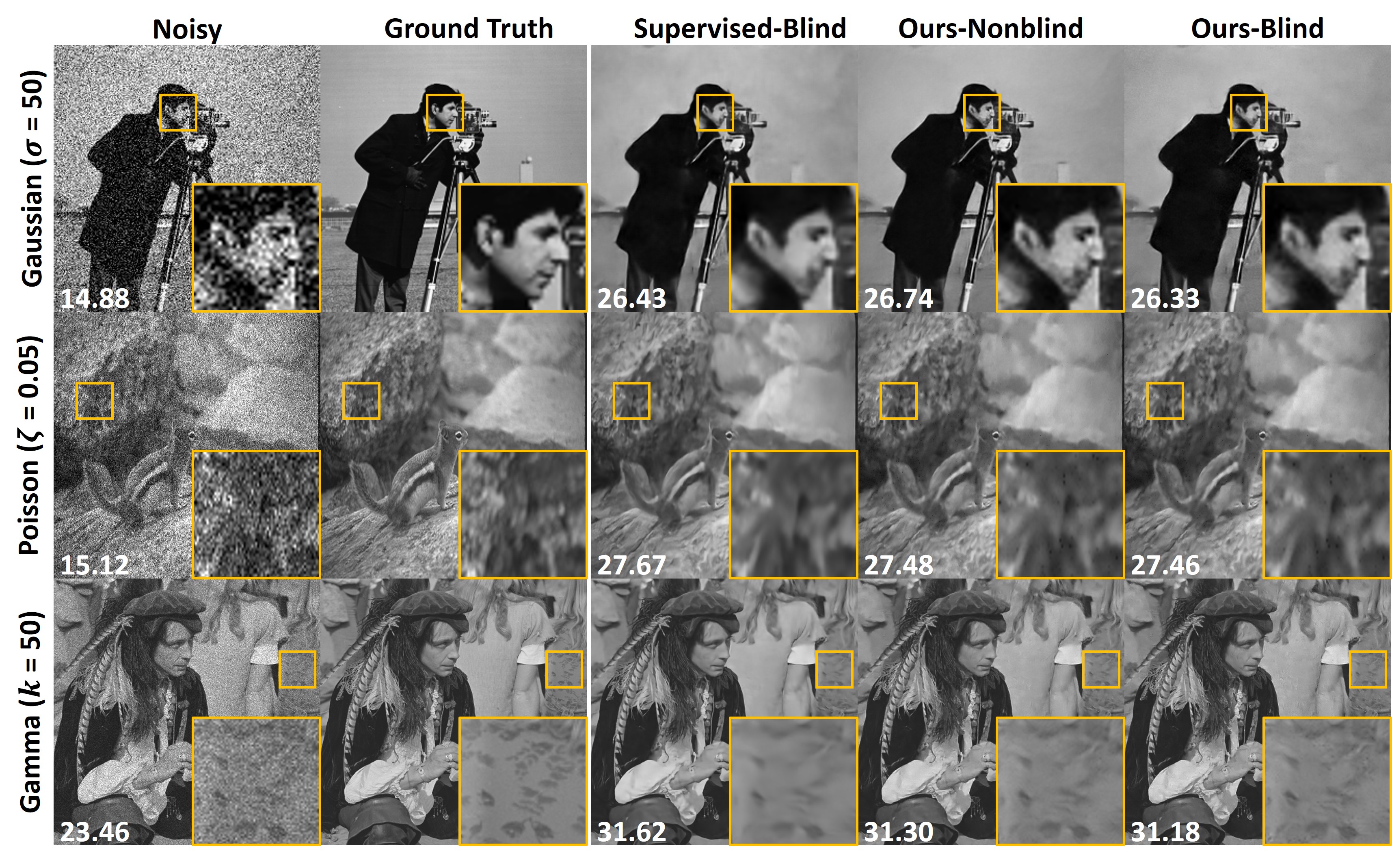}  
	\caption{Qualitative comparison for gray-scale images using Set12 and BSD68 dataset when the noise parameters are unknown. Top : Gaussian noise with $\sigma$= 50. Middle: Poisson noise with $\zeta$ = 0.05. Bottom: Gamma noise with $k$ = 50. White numbers at the lower left part of the images indicate the PSNR values in dB. Note that our blind approach provides comparable results to our method with known parameters.}
	\label{fig:blind_gray}
\end{figure}

\begin{figure}[!t]
	\centering
	\includegraphics[width=1\linewidth]{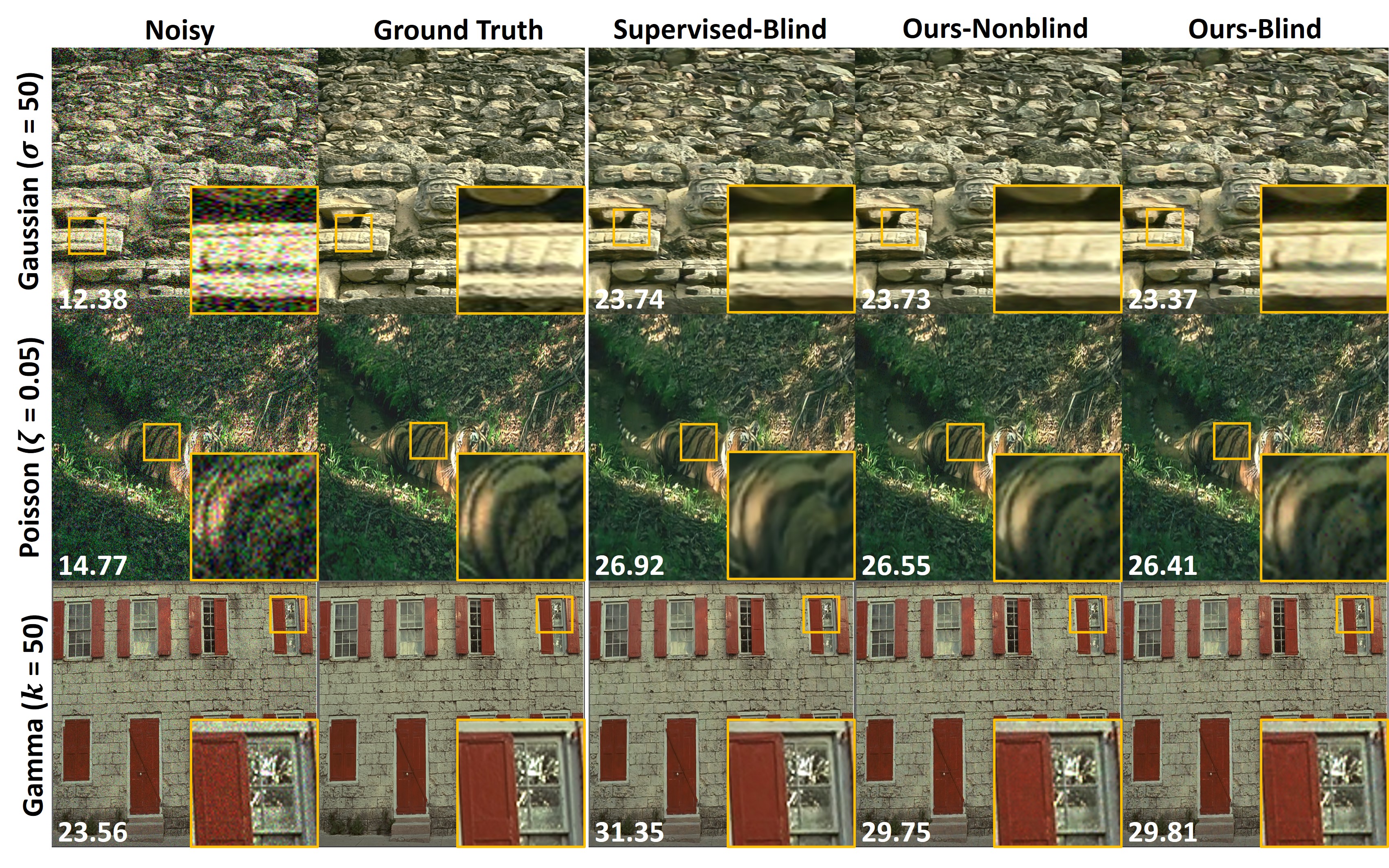}  
	\caption{Qualitative comparison for color-scale using Kodak and CBSD68 dataset when the noise parameters are unknown. Top : Gaussian noise with $\sigma$= 50. Middle: Poisson noise with $\zeta$ = 0.05. Bottom: Gamma noise with $k$ = 50. White numbers at the lower left part of the images indicate the PSNR values in dB. Noisy: noisy input. Note that our blind approach provides comparable results to our method with known parameters.}
	\label{fig:blind_color}
\end{figure}

Figs.~\ref{fig:blind_gray}, \ref{fig:blind_color} shows the visual results in case of unknown noise parameter for gray-scale and color image dataset, respectively. Note that our blind approach provides comparable results to our method with known parameters. Moreover, the results
are also comparable for supervised learning approach (supervised-blind) which is trained with
various noise level inputs and clean references.


\end{document}